\documentclass[%
 reprint,
 amsmath,amssymb,
 aps,
floatfix,
]{revtex4-2}

\usepackage{graphicx}
\usepackage{dcolumn}
\usepackage{bm}

\usepackage{xcolor}
 
\begin{document}

\preprint{APS/123-QED}

\title{A revision of the forces exerted in the Rayleigh regime by a tighlty focused optical vortex tweezer}

\author{Khalid Aloufi, Vasileios E. Lembessis and Omar M. Aldossary}
 \affiliation{Quantum Technology Group, Department of Physics and Astronomy, College of Science, King Saud
University, Riyadh 11451, Saudi Arabia}




\date{\today}

\begin{abstract}
Optical tweezers use light from a tightly focused laser beam to manipulate the motion of tiny particles. When the laser light is strongly focused, but still paraxial, its e/m field is characterized by a longitudinal component which is of magnitude comparable to the transverse ones and which has been ignored in the theoretical analysis of the tweezing forces. In our work we revise the calculations of the various components of the radiation pressure force, within the limits of Rayleigh regime or dipole approximation, in the case where a tiny particle interacts, in free space, with a circularly polarized optical vortex beam,  by taking into account this ignored field term. We show that this term is responsible for considerable modifications in the magnitude of the various components, moreover and also for the appearance of terms involving the coupling of the spin angular momentum (SAM) and the orbital angular momentum (OAM) of the photons of the vortex beam. We compare our findings with the ones taken ignoring the longitudinal field component.

\end{abstract}

\maketitle


\section{\label{sec:level1} Introduction}
The origin of optical tweezers is an experiment conducted in 1970 by Ashkin where
he succeeded in manipulating the motion of microscopic rubber beads by means
of a laser beam \cite{ashkin1970acceleration}. The optical tweezer was invented in 1986 by Ashkin and Steven Chu
when they managed to move, and at the same time trap, small dielectric beads \cite{ashkin1986observation}.
Since then we have witnessed a very broad range of applications of optical tweezers in biology, micromechanics, modelling of thermodynamic systems, collidal systems, and microchemistry \cite{jones2015optical}.

Optical tweezers are based on the forces that laser light exerts on tiny particles. Thus, the exact knowledge of these forces is prerequisite for any study involving the particles' dynamics in optical tweezers. When the dimensions of a particle are smaller than the wavelength of laser light then the motion of it can be described as a result of the action of radiation pressure force which is comprised of three distinct components namely: scattering force, dipole force and spin-curl force \cite{jones2015optical}.

Optical tweezers can operate at tight focussing conditions where the involved laser beams have a small beam waist. In this regime the electric and magnetic fields of light are characterized by a longitudinal component which has a magnitude comparable to the transverse electromagnetic fields components \cite{allen1999iv,lembessis2021chirality}. After the seminal work by Bliokh and Nori, \cite{bliokh2015transverse}, there is a growing interest on important contribution of this term in various physical effects including the mechanical effects of optical vortices on two-level atoms \cite{lembessis2021chirality},
 the modifications of the dressed states formalism in the case of two-level atoms \cite{lembessis2023two}, the optical properties of structured light beams \cite{koksal2023optical,koksal2022truncated,koksal2022hopf}, the realization of transverse angular momentum of light \cite{aiello2015transverse}, and its relation to the optical spin Hall effect \cite{banzer2013demonstration}. 

There are two important issues in the optical tweezers forces related with this field component. First it is usually ignored in the literature from the theoretical and mostly from numerical analysis. Second, when the light beam is in the form of a circularly polarized vortex, the longitudinal field component is responsible for the appearance of terms involving the coupling of the spin angular momentum (SAM) with the orbital angular momentum (OAM) of the beam's photons in the expressions of the radiation pressure force components \cite{lembessis2021chirality}.

This is exactly what we aim to do in the present work. We are going to show that this, so far ignored, longitudinal field component gives rise to considerable modifications in the magnitude of the radiation pressure forces while the coupling of photon SAM with OAM (which also arises from the logitudinal field component) brings in chiral effects in the forces expressions. 

In our work we will consider a tiny particle (whose dimensions are within the limits of Rayleigh regime or dipole approximation) irradiated, in free space, by a tightly focussed (but still paraxial) circularly polarized and we are going to find out the modified expressions for the various components of the radiation pressure force and we are going to compare our results with the "conventional" expressions for the radiation pressure force components which are in use by the community.

The structure of the paper is as follows. In Section II we give the analytical expressions for the three components of the radiation pressure force, at the beam focus $z=0$, as they have been calculated by taking into consideration the longitudinal field components. We present numerical results where we compare their magnitude with that of the radiation pressure force components when we ignore the longitudinal field component. In Section III we present our conclusions.

\section{\label{sec:level2}Optical tweezers forces}
The optical tweezers force is given by the following relation \cite{jones2015optical}:
\begin{equation}
  \mathbf{F}=\frac{1}{4} \alpha^{\prime} \nabla|\mathbf{E}\left(\mathbf{r}\right)|^{2}+\frac{1}{c} \sigma_{\text {ext }}\mathbf{S}\left(\mathbf{r}\right)-\frac{1}{2} \sigma_{\text {ext}} c \nabla \times \mathbf{s},
  \label{Eq:1}
\end{equation}
where is $\mathbf{S}=\frac{1}{2} \operatorname{Re}\left\{\mathbf{E} \times \mathbf{B}^{*}\right\} $ the time-averaged Poynting vector and $\mathbf{s}=i \frac{\varepsilon_{0}}{2 \omega}\left\{\mathbf{E} \times \mathbf{E}^{*}\right\}$ is the time-averaged spin density of the incoming wave, $c$  is the speed of light. The quantity $\sigma_{ext}=\frac{k_{0}\alpha^{\prime\prime}}{\varepsilon_{0}}$ is the extinction cross section, where $\alpha^{\prime}$ and $\alpha^{\prime\prime}$ are the real and imginary part of the complex polarizability respectively.\\\\
As we may see the radiation pressure force has three distinct and physically different components \cite{jones2015optical,ruffner2013comment,chen2011optical}:\\
(i) The gradient force $\mathbf{F}_{\text {grad}}\left(\mathbf{r}\right)=\frac{1}{4} \alpha^{\prime} \nabla|\mathbf{E}\left(\mathbf{r}\right)|^{2}$ which  is a conservative force, responsible for confinement in optical tweezers and it arises from the potential energy of a dipole in the electric field.\\
(ii) The scattering force $\quad \mathbf{F}_{ \text {scat}}\left(\mathbf{r}\right)=\frac{\sigma_{\mathrm{ext}}}{c} \mathbf{S}\left(\mathbf{r}\right)$ which is non-conservative and is responsible for the transfer of momentum and OAM from the field to the particle.\\
(iii) The spin-curl force, which is also non-conservative, $ \mathbf{F}_{ \mathrm{spin-curl}}\left(\mathbf{r}\right)=-\frac{1}{2} \sigma_{\mathrm{ext}} c \nabla \times \mathbf{s}\left(\mathbf{r}\right)$ which is also non-conservative and arises from polarization gradients in the electromagnetic field.\\\\
The two non-conservative components of the radiation pressure force draw their origin from the time averaged momentum, which in the free space can be separated into two parts, one which independent of the polarization and one which depends on circular polarization. The theoretical justification of this distinction rests Belifante's symmetrization of the canonincal stress energy tensor, which makes the results gauge independent. The spin-curl force is zero in homeogeneous fields and emerges when spatial inhomogeneities of polarization and intensity are present \cite{svak2018transverse}. This force has been recently experimentally measured in an experiment involving circularly polarized evanescent fields\cite{antognozzi2016direct}.

The electric field of a circularly polarized tightly focussed beam can be written as follows \cite{allen1999iv} :

\begin{equation} \label{Eq:2}
\mathbf{E}=\frac{1}{2}\biggl[\alpha E \vec{{x}}+\beta E \vec{{y}}+\frac{i}{k}\left(\alpha \frac{\partial E}{\partial x}+\beta \frac{\partial E}{\partial y}\right)\vec{{z}}\biggr] e^{(i k z-i \omega t)} +C.C., 
\end{equation}

where $\alpha$ and $\beta$ are in general complex constants with modulus $|\alpha|=|\beta|=1/\sqrt{2}$. Due to the cylindrical symmetry of optical vortex beams it is convenient to write the function $E$ in cylindrical polar coordinates as follows: $\ E=u(\rho, \phi, z)\exp (i\Theta(\rho, \phi, z))$
  
We consider the case where our vortex beam operates in a Laguerre-Gaussian mode
where the functions $u(\rho, \theta, \phi)$ and $\Theta(\rho, \theta, \phi)$ are given by:
\begin{equation}
   \begin{split}
           u&=E_{0}\sqrt{\frac{p !}{(p !+|\ell| !)}}\left(\frac{\rho \sqrt{2}}{w(z)}\right)^{|\ell|} \exp \left[-\frac{\rho^{2}}{w^{2}(z)}\right] L_{p}^{|\ell|}\left[\frac{ \rho^{2}}{w^{2}(z)}\right], \\
 \Theta&=\ell \phi-(2 p+|\ell|+1) \arctan \left(\frac{z}{z_{R}}\right)+\frac{k \rho^{2} z}{2\left(z^{2}+z_{R}^{2}\right)}.   
\end{split} 
\label{Eq:3}
\end{equation}
In the Eq.(\ref{Eq:3}) $\ell$ is the winding number, $p$ the radial index, $L_{p}^{|l|}$ is the associated Laguerre polynomial, $w(z)=w_{0} \sqrt{1+z^{2} / z_{R}^{2}}$ ,$w_{0}$  the beam waist, $z_{R}=\pi w_{0}^2 /  \lambda$ the Rayleigh range of the beam and $E_{0}$ the amplitude associated with the mode and is related to the beam power via the equation $E_{0}=\sqrt{\frac{4P}{\pi\epsilon_{0}c^{3}k^{2}w_{0}^{2}}}$ with $k=2\pi/\lambda$ being the wavenumber \cite{koksal2023optical}. 
With this notation the vortex electric field can be written as
\begin{align}
\begin{split}\label{Eq:4}
    \vec{E}=\frac{1}{2}\biggl\{\alpha u \vec{x}+\beta u\vec{y} +\frac{1}{k}\biggl[i\left(\alpha \frac{\partial u}{\partial x}+\beta \frac{\partial u}{\partial y}\right)\\
    -u\left(\alpha \frac{\partial \Theta}{\partial x}+\beta \frac{\partial \Theta}{\partial y}\right)\biggr] \vec{z}\biggr\} e^{i k z+i \Theta-i \omega t}+C.C.  
\end{split}  
\end{align}

\begin{figure}[t]
\frame{\includegraphics[width=0.96\linewidth,height=12cm]{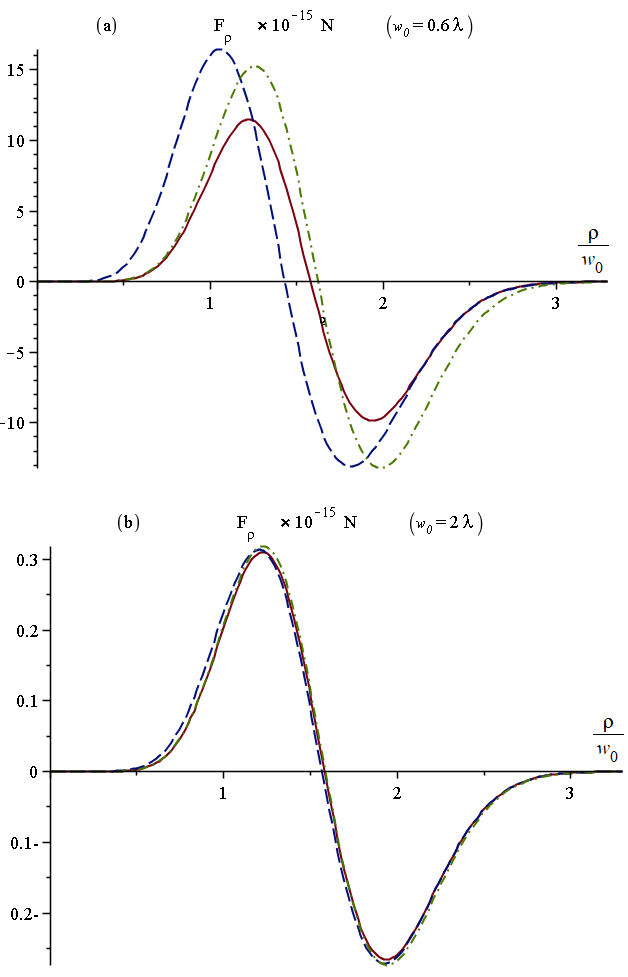}}
  \caption{The radial component of the dipole force, at the beam focus $(z=0)$, (a) up when the beam is tightly focused $(w_0=0.6 \lambda)$ (b) down when the beam is less focused to a beam waist $(w_0=2 \lambda)$.In both (a) and (b) the radial distances are in beam waist $w_0$ units. The solid line represents the conventional result where the longitudinal field components have not been taken into account, while the dashed line  represents the case where $\ell=+5$ and the dashed-dotted line represents the case $\ell=-5$}
\label{fig:1}
\end{figure}

\subsection{\label{level2} Dipole force}

\begin{figure}[t]
 \frame{\includegraphics[width=0.96\linewidth,height=12cm]{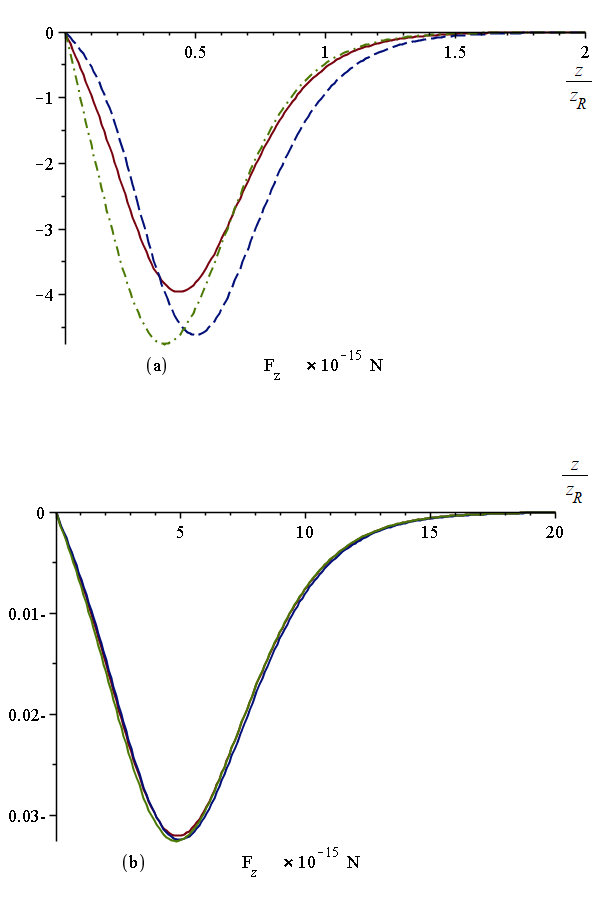}}
\caption{The axial component of the dipole force, at the radial distance $(\rho={w_0}\sqrt{\frac{|\ell|}{2}})$ (a) up when the beam is tightly focused $(w_0=0.6 \lambda)$ (b) down when the beam is not tightly focused $(w_0=2 \lambda)$.
  In both (a) and (b) axial distances are in Rayleigh range units $(z_R)$ . the sold line represents the conventional result where the
longitudinal field components have not been taken into
account, dash line  represents the case $\ell=+5$ and dash dot line represents the case $\ell=-5$ }
\label{fig:2}
\end{figure}

For the calculation of the dipole force we need the quantity $\nabla|\mathbf{E}\left(\mathbf{r}\right)|^{2}$. After calculations the expression for the dipole force, for a circularly polarized beam, is given by:
\begin{equation}
 \begin{split}\label{Eq:5}
&\mathbf{F}_{\text {grad}}\left(\mathbf{r}\right)=\frac{1}{4} \alpha^{\prime}\nabla|\mathbf{E}\left(\mathbf{r}\right)|^{2}\\
&= \frac{1}{4}\alpha^{\prime}\biggl\{\frac{1}{2} u \nabla u-\frac{i \sigma l}{4 k^{2}} \nabla\left(\frac{u}{\rho} \frac{\partial u}{\partial \rho}\right)+\frac{1}{8 k^{2}}\biggl[2\left(\frac{\partial u}{\partial \rho}\right) \nabla \frac{\partial u}{\partial \rho}\\
&+\nabla\left(u\left(\frac{\partial \Theta}{\partial \rho}\right)^{2}\right)+\frac{u l^{2}}{\rho^{2}}\left(\nabla u-\frac{u}{\rho^{}}\widehat{\rho}\right)\biggr]\biggr\},
\end{split}
\end{equation}
where $\sigma=(a*b^{*}-a^{*}b)$ is the spin of the light beam.
 The first term in the RHS is the "conventional" result which has been so far considered by the community. All the rest terms owe their existence to the longitudinal field component. The second term in the curly brackets, clearly shows a coupling between the SAM and OAM carried by the beam photons and vanishes when the beam is linearly polarized or when the beam photons carry no OAM (like an ordinary Gaussian beam). This terms exhibits chirality, in the sense that it reverses sign when either $\sigma$ or $\ell$ do.  Similarly the last term vanishes when there is no OAM in the beam. It is clear that the analytical work shows a considerable deviation from the conventional results. Let us see what the numerical work shows. 

 In all the numerical examples in this work we consider the following values for the involve parameters: an LG beam of wavelength $\lambda=633 \mathrm{~nm}$, a beam waist $w_{0}=0.65 \mu \mathrm{m}$, a power $P=10 \mathrm{~mW}$, a radial number $p=0$, a winding number $\ell=5$, and a counterclockwise circular polarization ($\sigma=i$). The beam irradiates a glass particle of relative dielectric permittivity $\varepsilon=2.25$ \cite{jones2015optical}.
 
 For the components of the dipole force, at beam focus $(z=0)$, we  get the plots shown in Fig.\ref{fig:1} and in Fig.\ref{fig:2} . In each plot we show three curves: a curve representing the "conventional" result where the longitudinal field components have not been taken into account (solid line), a curve which represents the case of positive winding number $+\ell$ (dashed line) and a curve for a negative winding number $-\ell$ (dashed-dotted line) .


In  Fig.\ref{fig:1}(a) we see that the magnitude of the maximum force increases considerably when we take into account the longitudinal component of the field with a slight shift of the maximum along the radial direction to the left in the case of +$\ell$ and to the right in the case  -$\ell$. The origin of this shifts in the maxima positions has to do with the fact that the the radial position at which the intensity maximizes is different when we take into account the longitudinal field components. Indeed if we ignore them we know that the radial distance at which the intensity maximizes is $w_{0}\sqrt{|\ell|/2}$ while in the present case this distance is modified by the existence of new terms in tightly focusing conditions. In  Fig.\ref{fig:1}(b) we see how the new results approximate the conventional results when we choose a less tighter focusing with a beam waist $w_{0}=2\lambda$. The same behavior is repeated when we calculate the axial component of the dipole force as it is shown in Figs.\ref{fig:2}(a)-(b). As we may see the plot shows the axial component as a function of the distance along the $z-$axis, keeping the radial distance equal to $w_{0}\sqrt{|\ell|/2}$. So the axial force is not plotted at the focus ($z=0$) as the radial component. The reason is that at the focus the axial force component is everywhere equal to zero along the radial direction. Our calculations have also shown that there is no component for dipole force along $\phi$ direction.






\subsection{\label{level2} Scattering force}
The scattering force is proportional to the time averaged Poynting vector and is given by the relation 
\begin{equation}
\begin{split}\label{Eq:6}
 &\mathbf{F}_{ \text{scat}}\left(\mathbf{r}\right)=\frac{\sigma_{\mathrm{ext}}}{c} \mathbf{S}\left(\mathbf{r}\right)\\
&= \frac{\varepsilon_{0}\sigma_{\mathrm{ext}}}{2c}\biggl[\mathrm{i}\sigma \left(E \nabla E^{*}-E^{*} \nabla E\right)+2  k|E|^{2} \hat{z}-\frac{\partial|E|^{2}}{\partial r} \hat{\boldsymbol{\phi}}\biggr].
\end{split}
\end{equation}
where, as we have said, $E=u e^{i \Theta}$. The first term in the square bracket of the second line has a zero contribution for linearly polarized light and is entirely due to the existence of the longitudinal field component. However new terms do exist, even in the case of a linearly polarized beam, in the axial and azimuthal scattering force components due to the new terms that exist in the expression of $|E|^{2}$. 

\begin{figure}[t]
 \frame{\includegraphics[width=0.96\linewidth,height=12cm]{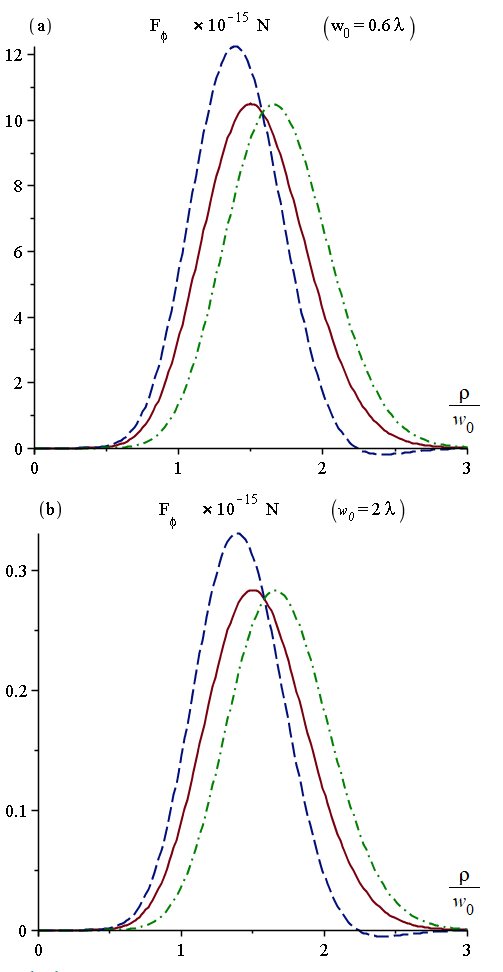}}
  \caption{The azimuthal component of the scattering force, at the beam focus $(z=0)$, (a) up when the beam is tightly focused $(w_0=0.6 \lambda)$ (b) down when the beam is not tightly focused $(w_0=2 \lambda)$.
  In both (a) and (b) the radial distances are in beam waist $w_0$ units. The sold line represents the conventional result where the longitudinal field components have not been taken into account and dash line  represents the case $\ell=+5$ }
\label{fig:3}
\end{figure}

We proceed to plot the scattering force components at $z=0$. At this position the radial force component is zero and the scattering force has only axial and azimuthal components. We use the same parameter values as in the case of the dipole force. In Figs.\ref{fig:3}(a)-(b) we present the azimuthal component of the scattering force. The solid one represents the force without taking into account  the longitudinal component of the field, the dashed one represents the same force but taking into account the longitudinal component of the electric field with positive $\ell$ and the dashed-dotted the magnitude of the same force when we consider a negative winding number $-\ell$. In the later case the sign of the force component becomes negative and this is the reason that we plot its magnitude. In  Fig.\ref{fig:3}(b) we see the same forces in the case of a relatively larger beam waist equal to $w_{0}=2\lambda$. As the beam waist increases (weaker focussing) the azimuthal scattering force component becomes weaker because the relevant beam intensity is smaller for given power and is similar in magnitude to the result obtained if we ignore the longitudinal field component.

\begin{figure}[t]
 \frame{\includegraphics[width=0.96\linewidth,height=12cm]{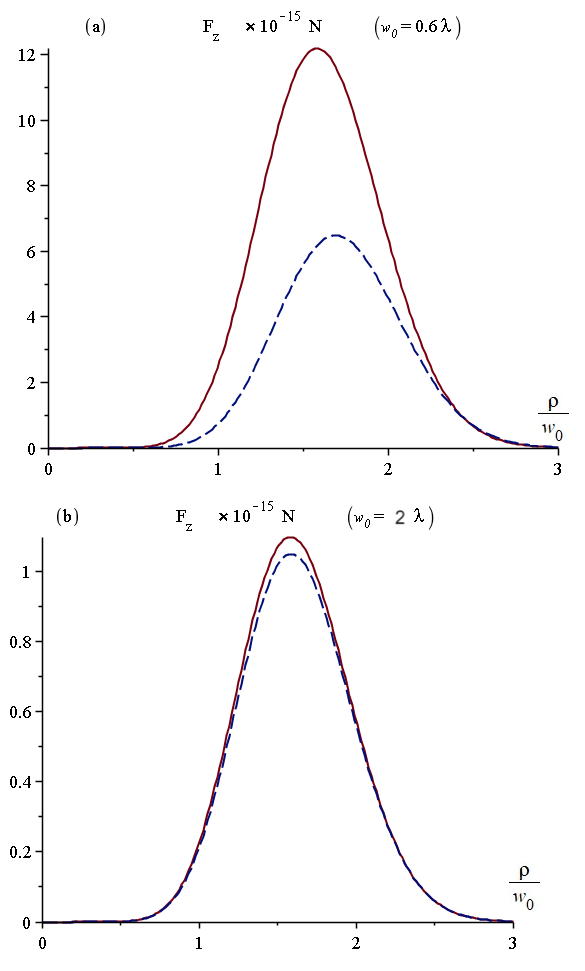}}
  \caption{The axial component of the scattering force at the beam focus $(z=0)$ (a) up when the beam is tightly focused $(w_0=0.6 \lambda)$ (b) down when the beam is not tightly focused $(w_0=2 \lambda)$. In both (a) and (b) the radial distances are in beam waist $w_0$ units. the sold line represents the conventional result where the longitudinal field components have not been taken into account and dash line  represents the case $\ell=+5$ }
\label{fig:4}
\end{figure}

In Figs.\ref{fig:4}(a)-(b) we present the axial component of the scattering force. The solid one represents the force without taking into account  the longitudinal component of the field, the dashed one represents the same force but taking into account the longitudinal component of the electric field with positive $\ell$ and the dashed-dotted the magnitude of the same force when we consider a negative winding number $-\ell$. In  Fig.\ref{fig:4}(b) we see the same forces in the case of a relatively larger beam waist equal to $w_{0}=2\lambda$. As the beam waist increases (weaker focusing) the azimuthal scattering force component becomes weaker because the relevant beam intensity is smaller for given power. The relevant relation between the conventional result (shown in solid line) and the new results not only remains but it becomes stronger (we need explanation here). In both Fig.\ref{fig:4}(a)-(b) we do not see the dashed-dotted curve which corresponds to the case of $-\ell$. The reason is that the axial scattering force does depend on $|\ell|$ and thus the two plots for $\pm\ell$ coincide at $z=0$ as it can be seen from the Eq.\ref{Eq:6}.


\subsection{\label{level3} Spin curl force}

\begin{figure}[t]
 \frame{\includegraphics[width=0.96\linewidth,height=12cm]{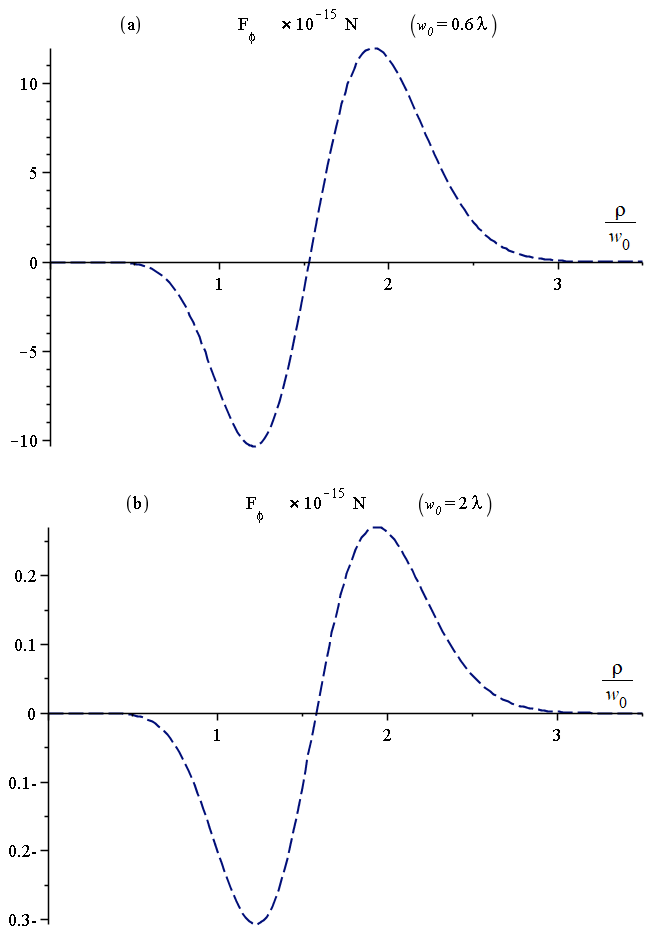}}
  \caption{The azimuthal component for the spin curl force, at the beam focus $(z=0)$, (a) up when the beam is tightly focused $(w_0=0.6 \lambda)$ (b) down when the beam is not tightly focused $(w_0=2 \lambda)$. In both (a) and (b) the radial distances  are in beam waist $w_0$ units the sold line represents the conventional result where the
longitudinal field components have not been taken into
account also it  represents the case $\ell=+5$  and the case $\ell=-5$}.
\label{fig:7}
\end{figure}

The spin curl force, as we have said, is given by $\mathbf{F}_{ \mathrm{spin}}\left(\mathbf{r}\right)=-\frac{1}{2} \sigma_{\mathrm{ext}} c \nabla \times \mathbf{s}\left(\mathbf{r}\right)$. The expression of the  spin curl force involves the calculation of the following quantity:
\begin{equation}
    \begin{split}\label{Eq:7}
\mathbf{F}_{ \mathrm{spin}}\left(\mathbf{r}\right)=  \frac{1}{16 k}&\alpha^{\prime\prime}\bigg\{\left[-i  \frac{\partial}{\partial z}\left(u \frac{\partial u}{\partial \rho}\right)-2 u\sigma \frac{\ell}{\rho} \frac{\partial u}{\partial z}\right] \widehat{\rho}\\
&+\sigma\left[ \frac{\partial}{\partial z}\left(u^2 \frac{\partial \Theta}{\partial\rho}\right)-2 u k \frac{\partial u}{\partial \rho}\right] \widehat{\phi}\\
& +{\frac{1}{\rho}}\left[i  \frac{\partial}{\partial \rho}\left(u \rho \frac{\partial u}{\partial \rho}\right)+2 u \sigma l \frac{\partial u}{\partial \rho}\right] \widehat{{z}}\biggr\}
    \end{split}
\end{equation}

\begin{figure}[tb]
 \frame{\includegraphics[width=0.96\linewidth,height=12cm]{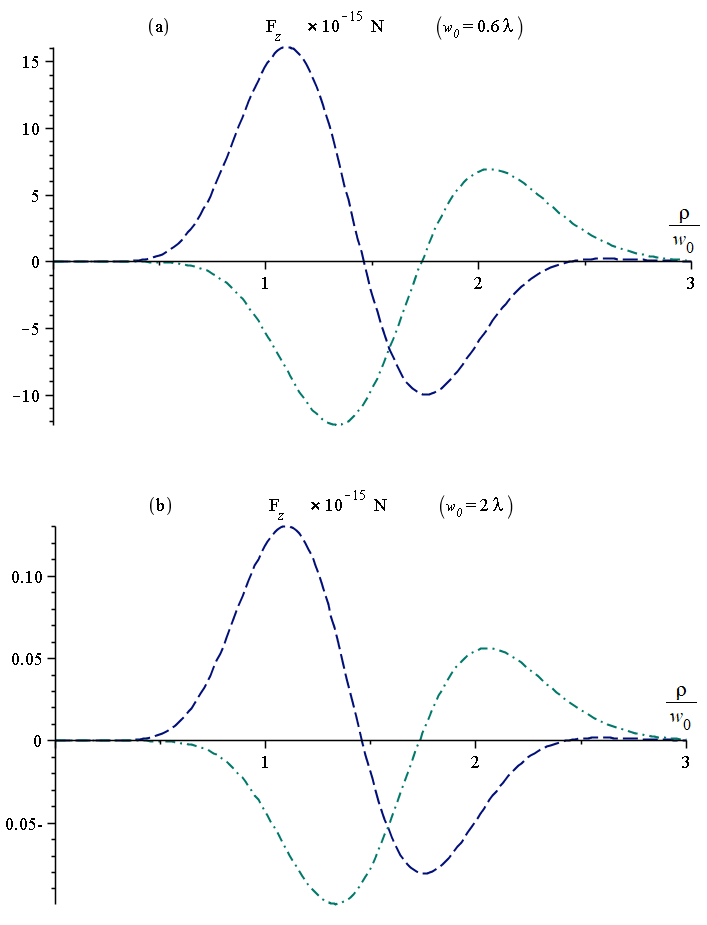}}
  \caption{The axial component of the spin curl force,at the beam focus $(z=0)$,  (a) up when the beam is tightly focused $(w_0=0.6 \lambda)$ (b) down when the beam is not tightly focused $(w_0=2 \lambda)$. In both (a) and (b) axial distances are in Rayleigh range units $(z_R)$. The dash line  represents the case $\ell=+5$ and the dash dot line represents the case $\ell=-5$ where the longitudinal field components have been taken into account.}
\label{fig:8}
\end{figure}

The magnitude of this force is very small compared to magnitude of dipole and scattering force and is normally ignored from the analysis. From Eq.\ref{Eq:7} we see that the radial component contains partial derivatives with respect to $z$ which means that the radial component vanishes at the focal plane $z=0$. The azimuthal component is proportional to $\sigma$ and thus vanishes for a linealry polarized beam. The axial component contains a term which couples the photon SAM with OAM, as well as the radial term but in the later case it vanishes on the focus plane. In Figs.\ref{fig:7}-\ref{fig:8} we present the azimuthal and axial components of the spin-curl force. It can be seen from the diagram that the inclusion of the longitudinal component of the field in the calculations does not affect the magnitude of the azimuthal component of the spin curl force. The force expression in beam foucs plane $(z=0)$ contains the absolute value of $\ell$ and therefore positive or negative has the same curve. \\

If the beam is not tightly focused we can see a significant decrease in this force with same behaviour.
Along the axial z-direction our calculations show that at $z=0$ there is no component in the conventional spin-curl force. Such a component arises only when we take into account the longitudinal field component since this field component actually creates a polarization gradient in the light field.
Our calculations have also shown that the spin-curl force has no radial at the beam focus plane.

\section{conclusions}
In our work we considered the interaction of a dielectric particle with a tightly focused circularly polarized LG beam. The size of the particle is smaller than the light wavelength and its interaction with the beam occurs in the free space. Due to strong focusing the electric and magnetic field of the beam have a longitudinal component with a magnitude comparable with the transverse ones. These components have been ignored from the analysis of the tweezing forces on the particle.  We calculated the three components of the radiation pressure force and we showed that there are considerable modifications in their magnitude when compared with the results obtained ignoring the longitudinal field component. Moreover, in the case of a circularly polarized vortex beam, we get new terms which arise from a coupling of the photon SAM and OAM. 
The longitudinal term is responsible for the appearance of a polarization gradient in the light field. As a result it is responsible for the appearance of a spin-curl force of magnitude comparable to the gradient and scattering forces which cannot be ignored, as it has so far been done in the literature where it was considered negligible. The spin curl force has a significant effect on the dynamics of a particle trapped in optical tweezers as has been demonstrated \cite{antognozzi2016direct}. The longitudinal term is present in any coherent tightly focused laser beam even in a Gaussian one. We have performed numerical and analytical work for such a beam and we have seen that the modifications are negligible. This shows clearly that the modifications in the expressions of the forces are a result of the coupling of the photon SAM and OAM which is present in the case of a tightly focused optical vortex. Our analysis concerns only the the calculation of forces in the Rayleigh regime. It is of great interest to investigate how the inclusion of the longitudinal field component will modify the expressions of the forces when the particle size is far larger than the optical wavelength (ray optics regime), or between these two regimes where we need to use the Lorentz-Mie theory or advanced numerical techniques.
\bibliography{ref}

\end{document}